\documentclass[12pt]{article}

\usepackage{epsfig}
\usepackage{graphicx}
\usepackage{a4}
\usepackage{latexsym}
\usepackage{cite}

\usepackage{pslatex}
\usepackage[latin1]{inputenc}
\usepackage[T1]{fontenc}
\newcommand{\ep}{\mbox{$\varepsilon$}}

\newcommand{\xo}{\mbox{$x_1^{\!{0}\!}$}}
\newcommand{\xt}{\mbox{$x_2^{\!{0}\!}$}}

\begin{document}
\setlength{\parskip}{0.2cm}
\setlength{\baselineskip}{0.55cm}
\begin{titlepage}
\begin{flushright} 
\hfill {\tt hep-ph/xxxxxxx}
\end{flushright} 
\begin{flushright} 
\hfill {\tt YITP-SB-07-24}
\end{flushright} 
\vspace{5mm} 
\begin{center} 
{\Large \bf 
Threshold corrections to rapidity distributions 
of $Z$ and $W^\pm$ bosons beyond N${}^2$LO at hadron colliders  
}\\
\end{center}

\vspace{10pt} 
\begin{center} 
{\bf 
V.~ Ravindran 
\footnote{ravindra@mri.ernet.in},   
J.~ Smith
\footnote{smith@max2.physics.sunysb.edu}.
}\\ 
\end{center}
\begin{center} 
{\it 
${}^1$Harish-Chandra Research Institute, 
 Chhatnag Road, Jhunsi, Allahabad, India,\\
${}^2$C.N. Yang Institute for Theoretical Physics,
Stony Brook University, Stony Brook, NY~11794-3840 USA.
} 
\end{center}
 
\vspace{10pt} 
\begin{center}
{\bf ABSTRACT} 
\end{center} 
Threshold enhanced perturbative QCD corrections to rapidity 
distributions of $Z$ and $W^\pm$ bosons at hadron colliders are presented  
using the Sudakov resummed cross sections at N${}^3$LO level.  We have used
renormalisation group invariance and the mass factorisation theorem 
that these hard scattering cross sections satisfy 
to construct the QCD amplitudes. We show that these higher order 
threshold QCD corrections stabilise the theoretical predictions 
for vector boson production at the LHC under 
variations of both renormalisation and factorisation scales. 
 
\vskip12pt 
\vskip 0.3 cm

\begin{center}
{\bf\large{\it This paper is dedicated to the memory of 
W.L.G.A.M. van Neerven.}}
\end{center}


\end{titlepage}
Recent theoretical advances in the computations of higher order 
radiative corrections in perturbative Quantum Chromodynamics (pQCD) 
have lead to extremely accurate predictions for several important observables  
needed for physics studies at the 
Tevatron collider in Fermilab as well as at the upcoming Large Hadron 
Collider (LHC) in CERN  \cite{Dittmar:2005ed}.
The Drell-Yan (DY) production of di-leptons \cite{Drell:1970wh},
which is one of the dominant production processes at hadron colliders, can 
be used to precisely calibrate the experimental detectors.
In addition, the DY process provides precise measurements of various
standard model parameters through measurements of the rapidity distributions 
of Z bosons \cite{Affolder:2000rx} and charge asymmetries of leptons 
coming from W boson decays \cite{Abe:1998rv}.
Possible excess events in di-lepton invariant mass distributions 
can point to physics beyond the standard model, such as $R$-parity violating
supersymmetric models, models with Z${}'$, 
or with contact interactions \cite{Affolder:2001ha}
and gravity mediated models (see \cite{Mathews:2004xp} for recent update).
The precise measurements of $Z$ and $W$ boson production 
cross sections, various distributions and asymmetries by both the D0 and 
CDF collaborations \cite{Patwa:2006rd} at the Fermilab Tevatron,
where $\sqrt{S} = $ 1.96 TeV, have 
already provided stringent tests of the standard model. These have already
played an important role in bounding the mass of the Higgs boson. 
Similar measurements at the LHC will provide even more stringent tests 
due to the increase in the number of events at $\sqrt{S}= $ 14 TeV.

The total cross sections for the $Z$ and $W^\pm$ production 
are known in pQCD up to next-to-next-to-leading order (N${}^2$LO) 
\cite{Kubar-Andre:1978uy,Altarelli:1978id,Humpert:1980uv,
Matsuura:1987wt, Matsuura:1988sm,Hamberg:1990np, Harlander:2002wh}.
Resummation programs for the threshold corrections 
to the total cross sections for DY production of di-leptons are also known  
\cite{Sterman:1986aj,Catani:1989ne}
(see also \cite{Kodaira:1981nh}) and one can
consult \cite{Vogt:2000ci,Catani:2003zt} for 
next-to-next-to-leading logarithmic (N${}^2$LL) resummation results so it is
straightforward to study the threshold effects in $Z$ and $W^\pm$ production
at the cross section level. 
Recent QCD results at the three loop level 
\cite{Moch:2004pa,Vogt:2004mw,Moch:2005id,Moch:2005tm,Vermaseren:2005qc,
Moch:2005ba, Blumlein:2004xt} have lead to predictions for  
the resummation up to N${}^3$LL \cite{Moch:2005ky,
Laenen:2005uz,Idilbi:2005ni,Ravindran:2006cg}. 
Notice that the fixed order partial-soft-plus-virtual N${}^3$LO corrections
\cite{Moch:2005ky,Ravindran:2006cg} 
to the Higgs and DY production show the reliability of the perturbation
theory results and demonstrate stability against the variations of 
renormalization and mass factorization scales.
Exact results up to N${}^2$LO are also available for less inclusive 
observables for di-lepton \cite{Anastasiou:2003yy}, $Z$ and $W^\pm$
\cite{Anastasiou:2003ds} 
(see \cite{Melnikov:2006di,Melnikov:2006kv,Cafarella:2007}) production. 
Recently the dominant QCD threshold corrections to the rapidity distribution 
of di-leptons in the DY process at N${}^3$LO have been obtained 
in \cite{Ravindran:2006bu}.  It was found that these corrections are 
indeed small and reduce the scale uncertainties significantly
making the predictions more reliable. 
The fixed order results as well as the resummed results reveal very  
interesting structures in the perturbative QCD series (see, 
\cite{Blumlein:2000wh,Blumlein:2006pj,
Blumlein:2005im,Dokshitzer:2005bf,Friot:2007fd,Ravindran:2005vv}).

The hard scattering cross sections computed using the QCD improved 
parton model
are often sensitive to variations in the renormalisation and factorisation 
scales usually denoted by $\mu_R$ and $\mu_F$ respectively.  The former 
originate from ultraviolet renormalisation while the latters originate
in the mass factorisation of collinear singularities.
In addition to the scale uncertainties the fixed order computations
suffer from the presence of various large logarithms which arise in some 
kinematical regions. These regions are often important from the experimental
point of view and these large lorgarithms, which spoil the standard 
perturbative predictions, should be resummed in a closed form.
For instance resummation formulae supplemented with fixed order 
results can predict the dominant higher order threshold corrections to 
various observables.  These threshold corrections are large when the 
fluxes of the incoming partons are large, which occurs at large partonic 
energies.

In \cite{Ravindran:2005vv} we computed the soft distribution functions that 
resum the soft gluons coming from real gluon emission processes in
DY production and Higgs production and also found that they are related by
the colour factor $C_A/C_F$,   
see also \cite{Ravindran:2004mb} and \cite{Ravindran:2003um}.
Using the soft distribution functions extracted
from DY, and the form factor of the Yukawa coupling of Higgs to
bottom quarks, we predicted 
the soft-plus-virtual (sv) parts of the Higgs production cross section
through bottom quark annihilation beyond N${}^2$LO
with the same accuracy that the DY process and the gluon fusion to
Higgs process are known \cite{Moch:2005ky,Ravindran:2006cg}.
This approach was then successfully applied in \cite{Blumlein:2006pj} to  
Higgs decay to bottom quarks and hadroproduction 
in $e^+e^-$ annihilation.  Since our results
in \cite{Ravindran:2005vv,Ravindran:2006cg,Blumlein:2006pj} are  
related to that of the standard threshold resummation, we could 
determine \cite{Ravindran:2005vv} the threshold exponents $D_i^I$ up to 
three loop level for DY and Higgs production using our resummed soft 
distribution functions and the quantities $B_i^I$ for deep inelastic 
scattering, Higgs decay and the hadroproduction of Higgs bosons.  
In \cite{Ravindran:2006bu}  we extended this approach 
to include $x_F$ and rapidity differential cross sections 
for di-lepton pairs in DY production and for Higgs bosons in Higgs 
production processes.  In this paper we apply these same methods
to study the effects of the dominant threshold corrections at N${}^3$LO 
to the rapidity distributions of the $Z$ and $W^\pm$ bosons 
in hadron-hadron collisions.

In \cite{Ravindran:2006bu} we formulated a framework to resum the dominant 
soft gluon contributions coming from the threshold region 
to the $x_F$ and rapidity distributions of DY di-lepton pairs and 
Higgs bosons 
at hadron colliders in the $z_i$ $(i=1,2)$ space of the kinematic variables.
We recapitulate the main points here to make this paper more understandable.
The threshold region corresponds to $z_i \rightarrow 1$ and in this region
all the partonic cross sections are symmetric in $z_1 \leftrightarrow z_2$.
To obtain the resummed result, we used renormalisation group (RG) invariance, 
mass factorisation and Sudakov resummation of QCD amplitudes.  
Using the resummed results in $z_i$ space 
we predicted the sv parts (also called threshold corrections) 
of the dominant partonic $x_F$ and rapidity distributions beyond N${}^2$LO.  
We follow the similar approach here to obtain the dominant threshold 
corrections at N${}^3$LO level for the rapidity distributions of $Z$ 
and $W^\pm$ bosons at both the LHC and Tevatron energies. 
See \cite{laenensterman} for an early reference where the resummation for DY 
differential distributions at rapidity $Y=0$ (or $x_F=0$) was considered.

The differential cross section for producing a vector boson 
can be expressed as:
\begin{eqnarray}
{d^2 \sigma^{~J} \over dq^2 dy } =
\sigma^{~J}_{\rm Born}(x_1^0,x_2^0,q^2)\, W^I(x_1^0,x_2^0,q^2) ~,
\end{eqnarray}
where $q$ is the four-momentum of the vector boson. In our case $q^2=M_J^2$
where $J=Z,W^\pm$ but for convenience we use $q^2$ for most of this paper.
Later we will present plots for 
$ d^2 \sigma^{~J} / dq dy $, where $q$ now represents 
$\sqrt{q^2}= M_Z$ or $M_W$ for $Z$ and $W^\pm$ respectively.
Our normalisation is  
$W^I_{\rm Born}(x_1^0,x_2^0,q^2)=\delta(1-x_1^0)\delta(1-x_2^0)$.
The superscript $I$ represents light-quarks ($q$), gluons ($g$)
and heavy quarks ($b$) but we only need the first case here so $I=q$
for the rest of the paper.
The $x^0_{i}~(i=1,2)$ are related to $q^2$, the scaling variable 
$\tau=q^2/S$, and the rapidity $y$ of the vector boson $J$:  
\begin{eqnarray}
y={1 \over 2}
\log\left({p_2 \cdot q \over p_1 \cdot q}\right) 
={1 \over 2 } \log\left({x_1^0 \over x_2^0}\right), 
\quad \quad \quad \tau=x_1^0 x_2^0 ~.
\end{eqnarray}
Here $S=(p_1+p_2)^2$ is the square of the hadronic center of mass energy and 
$p_i$ are the momenta of incoming hadrons $P_i~(i=1,2)$.

In the QCD improved parton model, the function $W^I(x_1^0,x_2^0,q^2)$ 
can be expressed in terms of the fitted parton distribution functions
(PDFs) appropriately convoluted with perturbatively calculable 
partonic differential cross sections
denoted by $\Delta^I_{d,ab}$ as follows
\begin{eqnarray}
W^I(x_1^0,x_2^0,q^2) &=&\sum_{a,b=q,\overline q,g} 
\int_0^1 dx_1 \int_0^1 dx_2~ {\cal H}^I_{ab}(x_1,x_2,\mu_F^2) 
\nonumber\\[2ex]
&&\times \int_0^1 dz_1 \int_0^1 dz_2
~\delta(x_1^0-x_1 z_1)~ \delta(x_2^0-x_2 z_2)~ 
\Delta^I_{d,ab} (z_1,z_2,q^2,\mu_F^2,\mu_R^2) \,,
\end{eqnarray}
where the subscript $d$ denotes the particular differential distribution
one is studying ($y$, $x_F$ etc).
Here $\mu_R$ is the renormalisation scale and $\mu_F$ the factorisation scale.
The function ${\cal H}^I_{ab}(x_1,x_2,\mu_F^2)$ is the product of PDFs
$f_a(x_1,\mu_F^2)$ and $f_b(x_2,\mu_F^2)$ renormalised at the
factorisation scale $\mu_F$.  That is,
\begin{eqnarray}
{\cal H}^q_{ab}(x_1,x_2,\mu_F^2)&=& 
f^{P_1}_a(x_1,\mu_F^2)~ f^{P_2}_b(x_2,\mu_F^2)\,,
\end{eqnarray}
with $x_i~(i=1,2)$ the momentum fractions of the partons in the 
incoming hadrons.

The partonic cross sections can be expressed in terms of soft and hard parts.
The soft parts come from the soft gluons that appear in real emission as 
well as in the virtual processes.  The infra-red safe contributions from 
the soft gluons can be obtained by adding the soft parts of
the differential cross sections with the ultraviolet renormalised 
virtual contributions and performing mass factorisation using appropriate 
counter terms.  These combinations are called the "soft-plus-virtual" (sv) 
parts of the differential cross sections. 
Hence we write
\begin{eqnarray}
\Delta^I_{d,ab} (z_1,z_2,q^2,\mu_F^2,\mu_R^2) = 
\Delta^{{\rm hard}}_{I,ab}(z_1,z_2,q^2,\mu_F^2,\mu_R^2)
+\delta_{a\overline b} \Delta^{\rm sv}_{~d,I}(z_1,z_2,q^2,\mu_F^2,\mu_R^2), 
\quad \quad \quad I=q \,.
\end{eqnarray}
The hard parts of the differential cross sections 
$\Delta^{\rm hard}_{I,ab}(z_1,z_2,q^2,\mu_F^2,\mu_R^2)$  
can be obtained by the standard procedure(see 
\cite{Rijken:1994sh,Mathews:2004xp}).
The sv parts of the differential cross sections
are obtained using the method discussed in the \cite{Ravindran:2006bu}
so that 
\begin{eqnarray}
\Delta^{\rm sv}_{~d,I}(z_1,z_2,q^2,\mu_R^2,\mu_F^2)={\cal C} \exp
\Bigg({\Psi^I_d(q^2,\mu_R^2,\mu_F^2,z_1,z_2,\ep)}\Bigg)\Bigg|_{\ep=0}\,,
\label{master}
\end{eqnarray}
where the $\Psi^I_d(q^2,\mu_R^2,\mu_F^2,z_1,z_2,\ep)$ are finite 
distributions computed in $4+\ep$ dimensions and they take the form
\begin{eqnarray}
\Psi^I_d(q^2,\mu_R^2,\mu_F^2,z_1,z_2,\ep)&=&
\Bigg(
\ln \big|\hat F^I(\hat a_s,Q^2,\mu^2,\ep)\big|^2
\Bigg)
\delta(1-z_1) \delta(1-z_2)
\nonumber\\[2ex]
&&+2~ \Phi^{~I}_d(\hat a_s,q^2,\mu^2,z_1,z_2,\ep)
- {\cal C}\ln \Gamma_{II}(\hat a_s,\mu^2,\mu_F^2,z_1,\ep)~ \delta(1-z_2)
\nonumber\\[2ex]
&&- {\cal C}\ln \Gamma_{II}(\hat a_s,\mu^2,\mu_F^2,z_2,\ep)~ \delta(1-z_1)
\,. 
\label{DYH}
\end{eqnarray}
The symbol "${\cal C}$" means convolution.  For 
example, ${\cal C}$ acting on the exponential
of a function $f(z_1,z_2)$ means the following expansion:
\begin{eqnarray}
{\cal C}e^{\displaystyle f(z_1,z_2) }&=& \delta(1-z_1)\delta(1-z_2)  + {1 \over 1!} f(z_1,z_2)
 +{1 \over 2!} f(z_1,z_2) \otimes f(z_1,z_2) 
\nonumber\\[2ex]
&& + {1 \over 3!} f(z_1,z_2) \otimes f(z_1,z_2) 
 \otimes f(z_1,z_2)
+ \cdot \cdot \cdot \,.
\end{eqnarray}
In the rest of the paper the function $f(z_1,z_2)$ 
is a distribution of the kind $\delta(1-z_j)$ or ${\cal D}_i(z_j)$, where
\begin{eqnarray}
{\cal D}_i(z_j)=\Bigg[{\ln^i(1-z_j) \over (1-z_j)}\Bigg]_+
\quad \quad \quad i=0,1,\cdot\cdot\cdot ,\quad {\rm and} \quad  j=1,2 \,,
\end{eqnarray}
and the symbol $\otimes$ means the "double" Mellin convolution 
with respect to the variables $z_1$ and $z_2$.
We drop all the regular functions that result from these convolutions
when defining the sv part of the cross sections. 
The $\hat F^I(\hat a_s,Q^2,\mu^2,\ep)$ are the standard form factors
coming from the purely virtual parts of the cross sections.
In the form factors, we have $Q^2=-M_J^2$.
The partonic cross sections depend on two scaling variables $z_1$ and $z_2$. 
The functions $\Phi^{~I}_d(\hat a_s,q^2,\mu^2,z_1,z_2,\ep)$ are
called the soft distribution functions.  
The unrenormalised (bare) strong coupling constant
$\hat a_s$ is defined as
\begin{eqnarray}
\hat a_s={\hat g^2_s \over 16 \pi^2}\,,
\end{eqnarray}
where $\hat g_s$ is the strong coupling constant which is dimensionless in
$n=4+\ep$ space time dimensions.  The scale $\mu$ comes from 
dimensional regularisation which makes the bare coupling constant $\hat g_s$
dimensionless in $n$ dimensions.
The bare coupling constant $\hat a_s$ is related to renormalised one by
the following relation:
\begin{eqnarray}
S_{\ep} \hat a_s = Z(\mu_R^2) a_s(\mu_R^2) 
\left(\mu^2 \over \mu_R^2\right)^{\ep \over 2}\,,
\label{renas}
\end{eqnarray}
where $S_{\ep}=\exp\left\{{\ep \over 2} [\gamma_E-\ln 4\pi]\right\}$
is the spherical factor characteristic of $n$-dimensional regularisation.
The renormalisation constant $Z(\mu_R^2)$ relates the bare coupling constant
$\hat a_s$ to the renormalised one $a_s(\mu_R^2)$.
They are both expressed in terms of the perturbatively calculable 
coefficients $\beta_i$ which are known up to four-loop level 
\cite{vanRitbergen:1997va,Czakon:2004bu} in terms of
the colour factors of SU(N) gauge group:
\begin{eqnarray}
C_A=N,\quad \quad \quad C_F={N^2-1 \over 2 N} , \quad \quad \quad
T_F={1 \over 2} \,.
\end{eqnarray}
Also we use $n_f$ for the number of active flavours.  

In dimensional regularisation, the bare form factors 
$\hat F^I(\hat a_s,Q^2,\mu^2,\ep)$ satisfy the following differential equation
\cite{Sudakov:1954sw,Mueller:1979ih,
Collins:1980ih,Sen:1981sd}.  
\begin{eqnarray}
Q^2{d \over dQ^2} \ln \hat {F^I}\left(\hat a_s,Q^2,\mu^2,\ep\right)&=&
{1 \over 2 }
\Bigg[K^I\left(\hat a_s,{\mu_R^2 \over \mu^2},\ep\right)
+ G^I\left(\hat a_s,{Q^2 \over \mu_R^2},{\mu_R^2 \over \mu^2},\ep\right)
\Bigg] \, .
\label{sud1}
\end{eqnarray}
The fact that the 
$\hat F^I(\hat a_s,Q^2,\mu^2,\ep)$ are  renormalisation group 
invariant and the functions $G^I$ are finite implies that
the $K^I$ terms can be expressed in terms of 
finite constants $A^I$, the so-called cusp anomalous dimensions and the
coefficients $\beta_i$.
The formal solution to the eqn.(\ref{sud1}), in dimensional regularisation, 
up to four-loop level is obtained in \cite{Moch:2005id,Ravindran:2005vv}.
The finite constants $G^{~I}_i(\ep)$ (see eqn.(19) of \cite{Ravindran:2006cg})
are also known \cite{Moch:2005tm} to the 
required accuracy in $\ep$.  
These constants  $G^{~I}_i(\ep)$  are expressed in terms of the functions 
$B_i^I$ and $f_i^I$. The $B_i^I$ are known up to order $a_s^3$  
through the three-loop anomalous dimensions (or splitting functions)
\cite {Moch:2004pa,Vogt:2004mw} and are found to be
flavour independent, that is $B_i^q=B_i^b$.
The constants $f_i^I$ are analogous to the cusp anomalous dimensions
$A_i^I$ that enter the form factors with $A_i^q=A_i^b$.  
It was first noticed in \cite{Ravindran:2004mb} that the single pole terms
in $\ep$ in the logarithms of the quark and gluon form factors up to two-loop 
level ($a_s^2$) can be predicted by the $C_F \rightarrow C_A$ substitution. 
The structure of single pole terms
of four-point amplitudes at the two-loop level can be found in 
\cite{Aybat:2006wq,Aybat:2006mz}.
The UV divergences present in the form factor are removed when the
bare coupling constant $\hat a_s$ undergoes renormalisation via
the eqn.(\ref{renas}).

The collinear singularities that arise due to the presence of
massless partons are
removed using the mass factorisation kernels $\Gamma(z_j,\mu_F^2,\ep)$ 
in the $\overline {\rm MS}$ scheme (see eqn.(\ref{DYH})).   
We suppress their dependence on $\hat a_s$ and $\mu^2$.
The factorisation kernels $\Gamma(z_j,\mu_F^2,\ep)$ satisfy the 
following renormalisation group equations:  
\begin{eqnarray}
\mu_F^2 {d \over d\mu_F^2}\Gamma(z_j,\mu_F^2,\ep)={1 \over 2}  P
                         \left(z_j,\mu_F^2\right)
                        \otimes \Gamma \left(z_j,\mu_F^2,\ep\right)
\quad\quad\quad j=1,2 \,,
\end{eqnarray}
where the $P(z_j,\mu_F^2)$ are the DGLAP matrix-valued 
splitting functions which are known up to three-loop 
level \cite{Moch:2004pa,Vogt:2004mw}:
\begin{eqnarray}
P(z_j,\mu_F^2)=
\sum_{i=1}^{\infty}a_s^i(\mu_F^2) P^{(i-1)}(z_j)\,.
\end{eqnarray}
The diagonal terms in the splitting functions $P^{(i)}(z_j)$ have the 
following structure
\begin{eqnarray}
P^{(i)}_{II}(z_j) = 2\Bigg[ B^I_{i+1} \delta(1-z_j)
         + A^I_{i+1} {\cal D}_0(z_j)\Bigg] + P_{{\rm reg},II}^{(i)}(z_j)\,,
\end{eqnarray}
where $P_{{\rm reg},II}^{(i)}(z_j)$ are regular when the argument approaches 
the kinematic limit (here $z_j \rightarrow 1$).
The RG equations can be solved by expanding them in powers of the
strong coupling constant.  Only the diagonal parts of the kernels contribute  
to the sv parts of the differential cross sections. 
We find the solutions contain only poles in $\ep$ in the 
$\overline{\rm MS}$ scheme:
\begin{eqnarray}
\Gamma(z_j,\mu_F^2,\ep)=\delta(1-z_j)+\sum_{i=1}^\infty \hat a_s^i
\left({\mu_F^2 \over \mu^2}\right)^{i {\ep \over 2}}S^i_{\ep}
\Gamma^{(i)}(z_j,\ep)\,.
\end{eqnarray}
An expansion for the $\Gamma^{(i)}(z_j,\ep)$ in negative powers of
$\ep$ up to four-loop level can be found in \cite{Ravindran:2005vv}.
The $\Gamma_{II}(\hat a_s,\mu^2,\mu_F^2,z_j,\ep)$ in eqn.(\ref{DYH})
is the diagonal element of $\Gamma(z_j,\mu_F^2,\ep)$.

From the eqn.(\ref{sud1}) and the fact that the 
$\Delta^{\rm sv}_{d,~I}$ are finite in the limit 
$\ep \rightarrow 0$ we obtain 
\begin{eqnarray}
q^2 {d \over dq^2}\Phi^{~I}_d(\hat a_s,q^2,\mu^2,z_1,z_2,\ep) =
{1 \over 2 }
\Bigg[\overline K^{~I}_d\left(\hat a_s,{\mu_R^2 \over \mu^2},z_1,z_2,\ep\right)
+ \overline G^{~I}_d\left(\hat a_s,{q^2 \over \mu_R^2},
{\mu_R^2 \over \mu^2},z_1,z_2,\ep\right)
\Bigg]\,,
\label{sud2}
\end{eqnarray}
where now the constants $\overline K^{~I}_d$ contain all the 
singular terms in $\ep$ and the $\overline G^{~I}_d$
are finite functions of $\ep$.  
The functions $\Phi^{~I}_d(\hat a_s,q^2,\mu^2,z_1,z_2,\ep)$ also satisfy
the renormalisation group equations:
\begin{eqnarray}
\mu_R^2 {d \over d\mu_R^2}\Phi^{~I}_d(\hat a_s,q^2,\mu^2,z_1,z_2,\ep)=0\,.
\end{eqnarray}
The $\Phi^{~I}_d(\hat a_s,q^2,\mu^2,z_1,z_2,\ep)$ should contain the 
correct poles to cancel the poles
coming from $\hat F^I$,$Z^I$ and $\Gamma_{II}$ in order to
make $\Delta^{\rm sv}_{~d,I}$ finite.  This requirement unambiguously 
determines all the poles of this distribution.  The solution to the Sudakov
differential equation for the soft distribution functions 
in eqn.(\ref{sud2}) can be
written as
\begin{eqnarray}
\Phi^{I}_d(\hat a_s,q^2,\mu^2,z_1,z_2,\ep) &=& 
\sum_{i=1}^\infty \hat a_s^i S_{\ep}^i 
\left({q^2 (1-z_1)(1-z_2) 
\over \mu^2}\right)^{i {\ep \over 2}} 
\left({(i~\ep)^2 \over 4(1-z_1) (1-z_2)} \right)
\hat \phi^{I,(i)}_d(\ep)\,,
\label{softsol}
\end{eqnarray}
where 
\begin{eqnarray}
\hat \phi^{~I,(i)}_d(\ep)=
{1 \over i \ep} \Bigg[ \overline K^{~I,(i)}_d(\ep) 
+ \overline {G}^{~I,(i)}_d(\ep)\Bigg]\,.
\end{eqnarray}
The constants $\overline K^{~I,(i)}_d(\ep)$
are expanded in powers of the 
bare coupling constant $\hat a_s$ as follows
\begin{eqnarray}
\overline K^I_d\left(\hat a_s,{\mu_R^2\over \mu^2},z_1,z_2,\ep\right)
=\delta(1-z_1)\delta(1-z_2) \sum_{i=1}^\infty \hat a_s^i
\left({\mu_R^2 \over \mu^2}\right)^{i {\ep \over 2}}S^i_{\ep}~
\overline K^{~I,(i)}_d(\ep) \,.
\end{eqnarray}
Using the RG equation for
$\overline K^I_d\left(\hat a_s,\mu_R^2/\mu^2,z_1,z_2,\ep\right)$,  
one finds that the constants $\overline K^{~I,(i)}_d(\ep)$ are identical 
to $\overline K^{~I,(i)}(\ep)$ given 
in \cite{Ravindran:2006cg}.
The constants $\overline {G}^{~I,(i)}_d(\ep)$ are related to the finite
boundary functions $\overline G^I_d(a_s(q^2),1,z_1,z_2,\ep)$. 
We define the $\overline {\cal G}_{d,i}^I(\ep)$ through the relation 
\begin{eqnarray}
\sum_{i=1}^\infty \hat a_s^i 
\left( {q^2 (1-z_1)(1-z_2) \over \mu^2}\right)^{i{\ep \over 2}} 
S^i_{\ep}
\overline G_d^{~I,(i)}(\ep)
&=&
\sum_{i=1}^\infty a_s^i\left(q^2 (1-z_1)(1-z_2)\right) 
\overline {\cal G}^{~I}_{d,i}(\ep)\,.
\label{Gbar1}
\end{eqnarray}
We obtain the $z_1,z_2$ independent constants
$\overline {\cal G}^{~I}_{d,i}(\ep)$ 
by demanding the finiteness of $\Delta^{\rm sv}_{~d,I}$ given in
eqn.(\ref{master}).   
Before setting $\ep=0$ in eqn.(\ref{master}), we expand 
$\Delta^{\rm sv}_{~d,I}$ as
\begin{eqnarray}
\Delta^{\rm sv}_{~d,I}(z_1,z_2,q^2,\mu_R^2,\mu_F^2,\ep)
=\sum_{i=0}^\infty a_s^i(\mu_R^2)
\Delta^{\rm sv,(i)}_{~d,I}(z_1,z_2,q^2,\mu_R^2,\mu_F^2,\ep)\,.
\end{eqnarray}
Using the above expansion and eqn.(\ref{DYH}) we determine 
these constants using the known information on the form factors, 
the mass factorisation kernels and the coefficient
functions $\Delta^{\rm sv,(i-1)}_{~d,I}$ expanded in powers of $\ep$.
The structure of the $G^{~I}_d(\ep)$ in the form factors 
involving the constants $f^I$ and $\beta_i$ was given 
in \cite{Ravindran:2006cg}.  
The constants $\overline {\cal G}^{~I}_{d,i}(\ep)$ in the soft 
distribution functions also have a similar structure:
\begin{eqnarray}
\overline {\cal G}^{~I}_{d,1}(\ep)&=&-f_1^I+
\sum_{k=1}^\infty \ep^k \overline {\cal G}^{~I,(k)}_{d,1}\,,
\nonumber\\[2ex]
\overline {\cal G}^{~I}_{d,2}(\ep)&=&-f_2^I
-2 \beta_0 \overline{\cal G}_{d,1}^{~I,(1)}
+\sum_{k=1}^\infty\ep^k  \overline {\cal G}^{~I,(k)}_{d,2}\,,
\nonumber\\[2ex]
\overline {\cal G}^{~I}_{d,3}(\ep)&=&-f_3^I
-2 \beta_1 \overline{\cal G}_{d,1}^{~I,(1)}
-2 \beta_0 \left(\overline{\cal G}_{d,2}^{~I,(1)}
+2 \beta_0 \overline{\cal G}_{d,1}^{~I,(2)}\right)
+\sum_{k=1}^\infty \ep^k \overline {\cal G}^{~I,(k)}_{d,3}\,,
\nonumber \\[2ex]
\overline {\cal G}^{~I}_{d,4}(\ep)&=&
-f_4^I
-2 \beta_2 \overline {\cal G}^{~I,(1)}_{d,1}
-2 \beta_1 \Big( \overline {\cal G}^{~I,(1)}_{d,2} 
+ 4 \beta_0 \overline {\cal G}^{~I,(2)}_{d,1} \Big)\,,
\nonumber\\[2ex]
&&-2 \beta_0 \Big(\overline {\cal G}^{~I,(1)}_{d,3}
+2 \beta_0 \overline {\cal G}^{~I,(2)}_{d,2}
                         +4 \beta_0^2 \overline {\cal G}^{~I,(3)}_{d,1}\Big)
        +\sum_{k=1}^\infty \ep^k \overline {\cal G}^{~I,(k)}_{d,4}\,.
\label{OGI}
\end{eqnarray}
The terms proportional to $\ep$ at every order in $\hat a_s$ are determined 
using the known $\sigma^J$ to order N${}^2$LO and the following identity:
\begin{eqnarray}
\int_0^1 
dx_1^0 \int_0^1 
dx_2^0 \left(x_1^0 x_2^0\right)^{N-1}
{d \sigma^{~J} \over d y}
=\int_0^1 d\tau~ \tau^{N-1} ~\sigma^{~J}\,.
\label{iden}
\end{eqnarray}
We find
\begin{eqnarray}
\overline{\cal G}^{~q,(1)}_{d,1}
&=&C_F~ \Big(- \zeta_2\Big) \,,
\nonumber\\[2ex]
\overline{\cal G}^{~q,(2)}_{d,1}
&=& C_F~ \Bigg({1 \over 3}  \zeta_3\Bigg)\,,
\nonumber\\[2ex]
\overline{\cal G}^{~q,(3)}_{d,1}
&=& C_F~ \Bigg({1 \over 80}  \zeta_2^2\Bigg)\,,
\nonumber\\[2ex]
\overline{\cal G}^{~q,(1)}_{d,2}
&=& C_F C_A~ \Bigg({2428 \over 81} -{67 \over 3} \zeta_2
              -4 \zeta_2^2 -{44 \over 3} \zeta_3\Bigg)
\nonumber\\[2ex]
&&             +C_F n_f~ \Bigg(-{328 \over 81} + {10 \over 3} \zeta_2
                +{8 \over 3} \zeta_3 \Bigg)\,.
\label{DIi}
\end{eqnarray}

Using the resummed result given in eqn.(\ref{master}), the
exponents $g_i^{~q}(\ep)$ (see \cite{Moch:2005tm}) and 
$\overline {\cal G}_{d,i}^{~q}(\ep)$, 
we can obtain the higher order sv 
contributions to the differential cross sections.  The available
exponents are 
\begin{eqnarray}
&g_1^{~q,j}~,~~~
\overline {\cal G}_{d,1}^{~q,(j)} \hspace{1.5cm} 
{\rm for}\hspace{1.5cm}  j={\rm all} \quad \,,
\nonumber\\[2ex]
&g_2^{~q,j}~,~~~
\overline {\cal G}_{d,2}^{~q,(j)} \hspace{1.5cm} 
{\rm for} \hspace{1.5cm} j=0,1 \quad \,,
\nonumber\\[2ex]
&g_3^{~q,j}~,~~~
\overline {\cal G}_{d,3}^{~q,(j)} \hspace{1.5cm} 
{\rm for} \hspace{1.5cm} j=0 \quad \,,
\nonumber
\end{eqnarray}
in addition to the known $\beta_i~(i=0,1,2,3)$ and the constants in the 
splitting functions $A^q_i,~ B^q_i~~(i=1,2,3)$ and $f^q_i~(i=1,2,3)$.  
The constants $g^{q,j}_2$ are known for $j=2,3$ also
(see \cite{Moch:2005id}).  
Using our approach we have obtained the exact 
$\Delta_{d,q}^{\rm sv,(i)}$ up to N${}^2$LO ($i=0,1,2$) \cite{Anastasiou:2003ds}.
The coefficient of the $\delta(1-z_1)\delta(1-z_2)$ part depends on
the constants $\overline {\cal G}^{q,(2)}_2,g^{q,1}_3,\overline 
{\cal G}^{q,(1)}_3$ which are still unknown for N${}^3$LO, so
we can only obtain {\it a partial result} for $\Delta_{d,q}^{\rm sv,(3)}$, 
i.e., a result without the $\delta(1-z_1)\delta(1-z_2)$ part can be 
computed from our formula given in eqn.(\ref{master}).  
We can also obtain a result  
to N${}^4$LO order where we can predict {\it partial} sv
contributions containing everything except 
the terms in ${\cal D}_0(z_i) \delta(1-z_j),{\cal D}_0(z_i) {\cal D}_0(z_j),
{\cal D}_1(z_i) \delta(1-z_j)$ and $\delta(1-z_1) \delta(1-z_2)$ for
the coefficient $\Delta_{d,q}^{\rm sv,(4)}$. 
The results are identical to those given in  
the Appendix B of \cite{Ravindran:2006bu} for $\mu_R^2=\mu_F^2=M_J^2$.   
The convolutions of distributions of the form 
${\cal D}_l(z_j) \otimes {\cal D}_m(z_j)$ for any arbitrary $l,m$
can be done using the general formulae given in \cite{Ravindran:2006cg} 
so we obtain $\Delta_{d,I}^{\rm sv,(i)}$ for $i=1,...,4$.
\begin{figure}[htb]
\vspace{1mm}
\centerline{
\epsfig{file=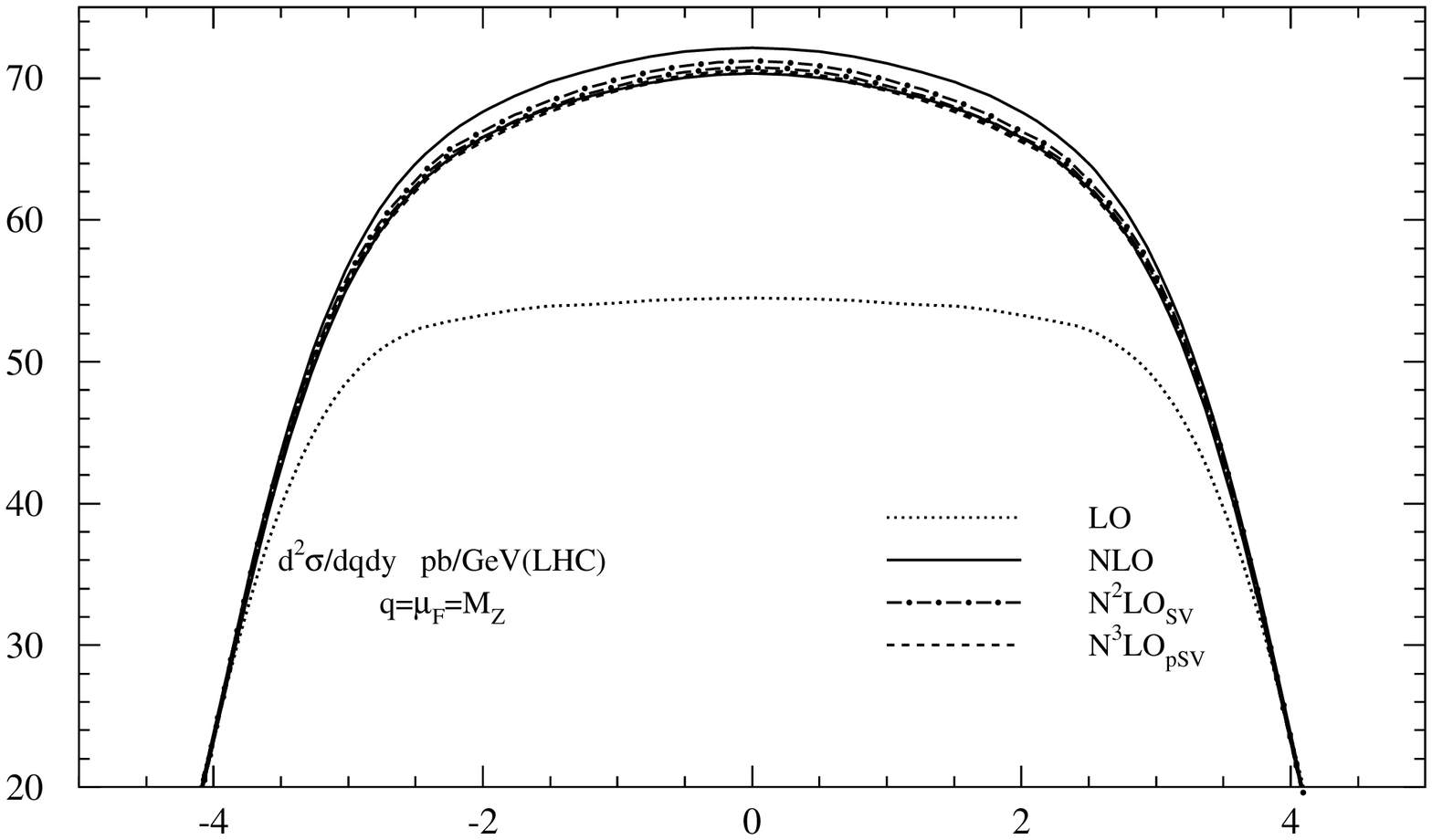,width=8cm,height=10cm,angle=0}
\epsfig{file=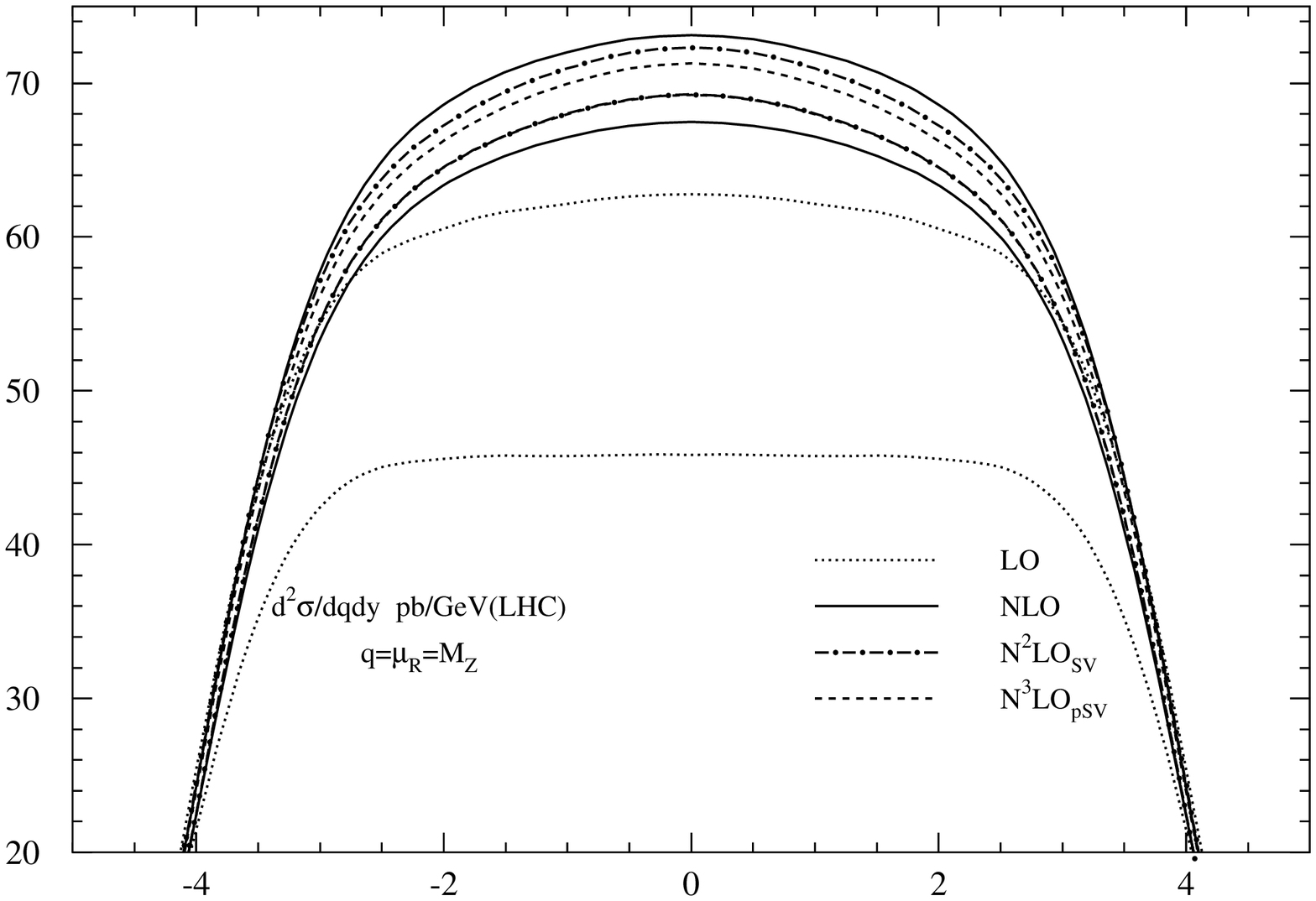,width=8cm,height=10cm,angle=0}
}
\caption{
Rapidity distributions for $Z$ boson production at the LHC,
and their $\mu=\mu_R$ (left panel) and $\mu=\mu_F$ (right panel) 
scale dependence (with $M^2_Z/2 < \mu^2 <2 M^2_Z$).  
The abbreviation "pSV" means partial-soft-plus-virtual.
}
\end{figure}
\begin{figure}[htb]
\vspace{1mm}
\centerline{
\epsfig{file=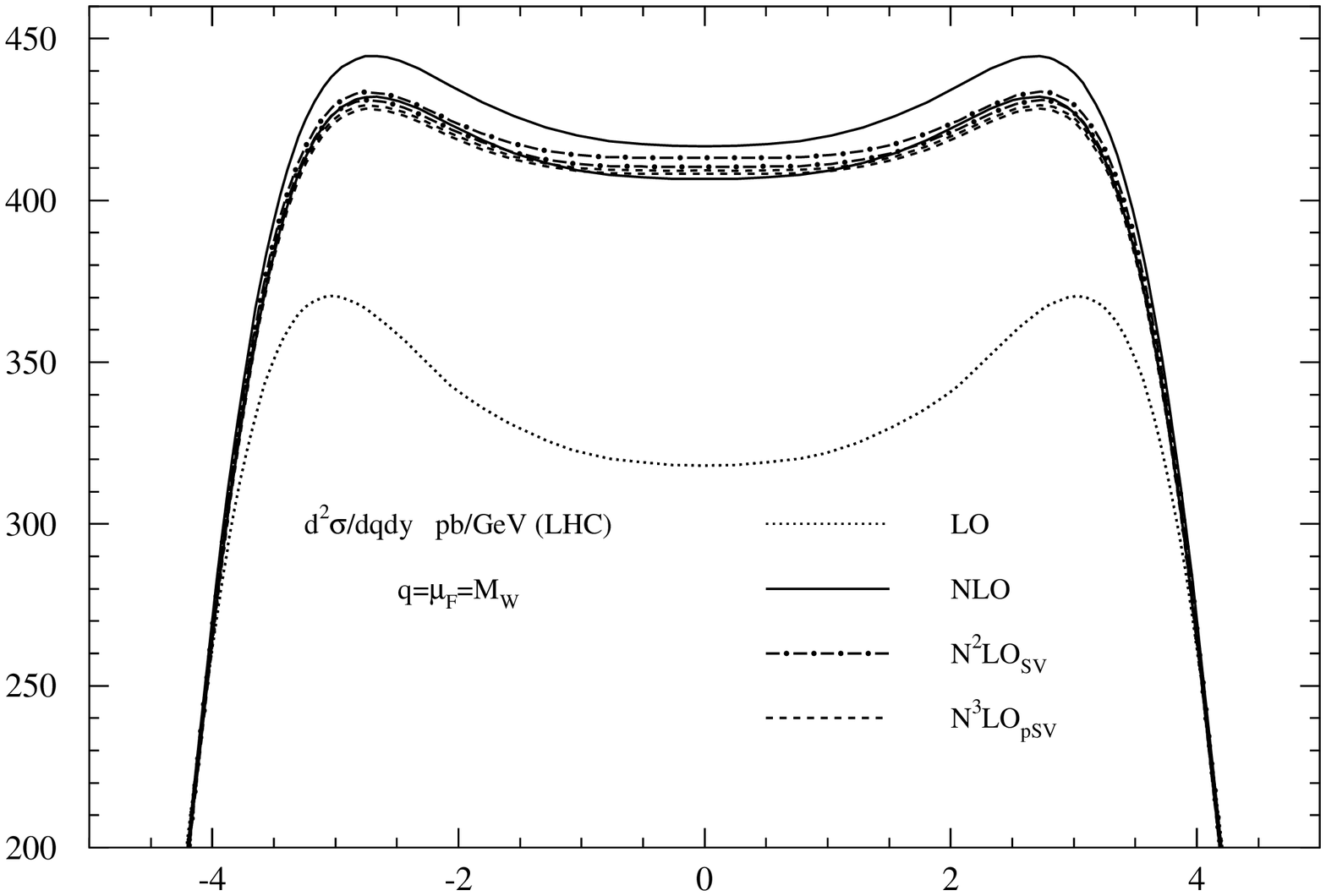,width=8cm,height=10cm,angle=0}
\epsfig{file=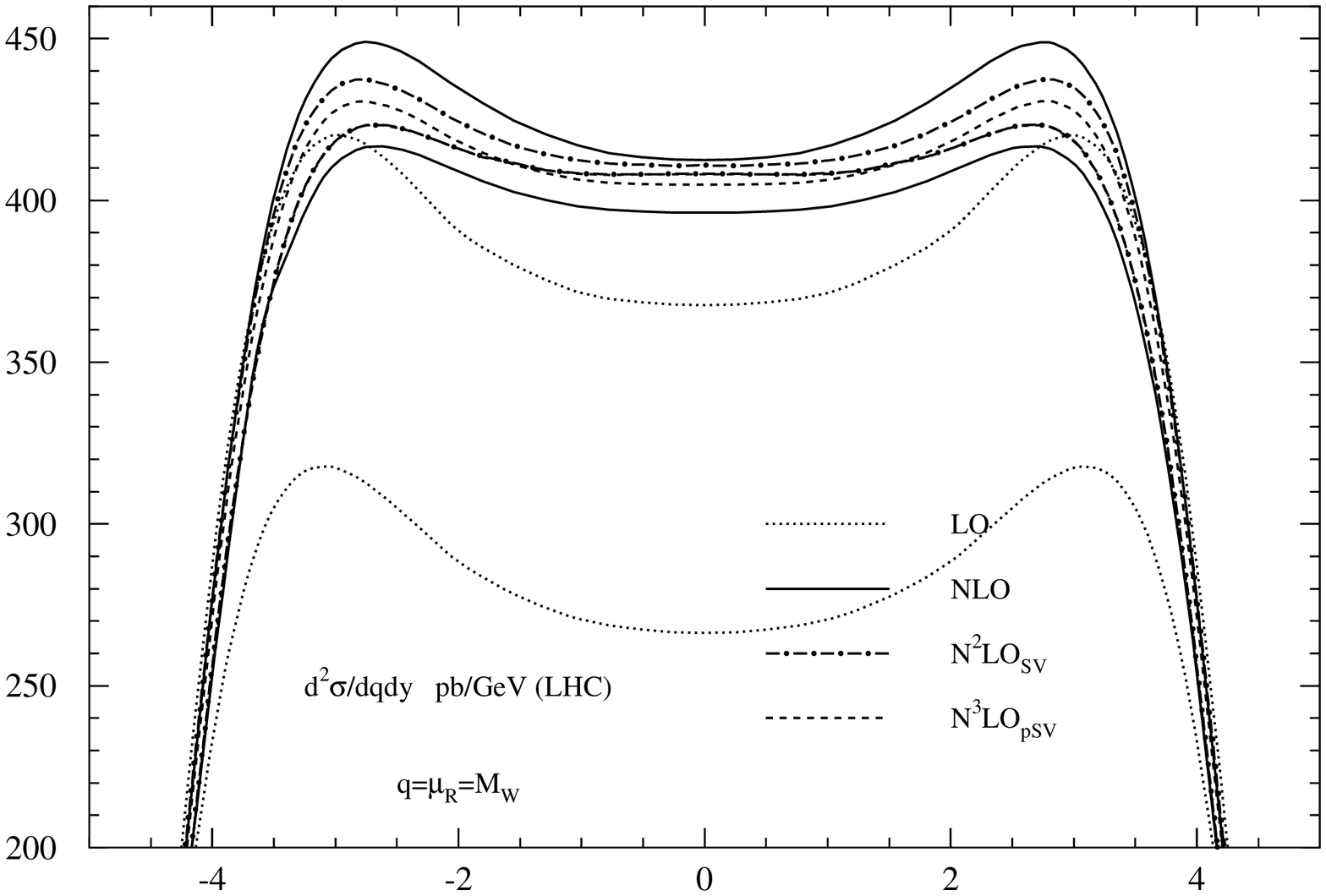,width=8cm,height=10cm,angle=0}
}
\caption{
Rapidity distributions for $W^+$ boson production at the LHC,
and their $\mu=\mu_R$ (left panel) and $\mu=\mu_F$ (right panel) 
scale dependence (with $M^2_W/2 < \mu^2 <2 M^2_W$).  
The abbreviation "pSV" means partial-soft-plus-virtual.
}
\end{figure}
The differential cross sections can be expanded in powers of
the strong coupling constant as 
\begin{eqnarray}
{d^2 \sigma^J\over  dq^2 dy} &=& \sum_{i=0}^\infty a_s^i 
~{d^2 \sigma^{J,(i)} \over dq^2 dy} \,.
\end{eqnarray}
We split the partonic cross section into hard and sv parts:
\begin{eqnarray}
{d^2 \sigma^{J,(i)} \over dq^2 dy} = 
{d^2 \sigma^{{\rm hard},J,(i)} \over dq^2 dy}
+{d^2 \sigma^{{\rm sv},J,(i)} \over dq^2 dy}\,,
\end{eqnarray}
\begin{eqnarray}
2 S~ {d^2 \sigma^{{\rm hard},J,(i)}\over dq^2 dy} &=&
\sum_{q} {\cal G}_{SM}^J 
\Bigg( D^{SM,(i)}_{q \overline q}(\xo,\xt,\mu_F^2)
      +D^{SM,(i)}_{q g}(\xo,\xt,\mu_F^2)
\nonumber\\[2ex]&&
      +D^{SM,(i)}_{g q}(\xo,\xt,\mu_F^2)\Bigg) \,.
\end{eqnarray}

The SM coefficients  $D^{SM,(i)}_{ab}(\xo,\xt,\mu_F^2)$ can be found in
\cite{Rijken:1994sh,Mathews:2004xp}.
The sv part of the partonic cross section can be expressed as
\begin{eqnarray}
2 S~ {d^2 \sigma^{{\rm sv},J,(i)} \over dq^2 dy} 
= \sum_{a,b=q,\overline q}{\cal G}_{SM}^{J}
\int_0^1~ dx_1~ \int_0^1~ dx_1~ {\cal H}^q_{ab}(x_1,x_2,\mu_F^2)~ 
\nonumber\\[2ex]
\times \int_0^1 dz_1 \int_0^1 dz_2 ~\delta(x_1^0-x_1 z_1) 
~\delta(x_2^0-x_2 z_2)~ \Delta_{d,q}^{\rm sv,(i)}
(z_1,z_2,q^2,\mu_F^2,\mu_R^2)\,.
\end{eqnarray}
The coefficients $\Delta_{d,ab}^{\rm sv,(i)}(z_1,z_2,q^2,\mu_F^2,\mu_R^2)$ are 
presented in the Appendix B of \cite{Ravindran:2006bu}, with the normalisation 
$\Delta_{y,ab}^{\rm sv,(0)}(z_1,z_2,q^2,\mu_F^2,\mu_R^2)=
\delta(1-z_1)\delta(1-z_2)$.
The functions ${\cal G}_{SM}^{~J}$ are given by
\begin{eqnarray}
{\cal G}_{SM}^{M_Z}&=&{4 \alpha^2 Q_q^2 \over 3 q^2}
+{4\alpha q^2 \Gamma_{Z\rightarrow l^+ l^-} \over
M_Z \left((q^2-M_Z^2)^2+M_Z^2 \Gamma_Z^2\right) c_w^2 s_w^2}
\Big((g_q^V)^2+(g_q^A)^2\Big)
\nonumber\\[2ex]
&&+{2\alpha^2 (1-4 s_w^2) (q^2 - M_Z^2) 
\over  3 \left((q^2-M_Z^2)^2+M_Z^2 \Gamma_Z^2\right)
c_w^2 s_w^2}
Q_q g_q^V
\,,
\nonumber\\[2ex]
\label{ew}
\\[2ex]
{\cal G}_{SM}^{M_W}&=&{\alpha q^2 \Gamma_{W \rightarrow l \overline \nu_l} 
\over 
\left(\left(q^2 -M_W^2\right)^2 + M_W^2 \Gamma_W^2\right) M_W s_w^2
} V_{ij}^2 \,.
\label{gr}
\end{eqnarray}
We now give results for $Z$ and $W^\pm$ production by choosing $q^2=M_Z^2$ 
and $q^2=M_W^2$ respectively. At these points the functions above contain
the standard electro-weak constants which can be found in 
\cite{Mathews:2004xp,Ravindran:2003um} and the
standard CKM matrix elements $V_{ij}$. 
We present our results as differential cross sections 
in rapidity for these fixed $q^2$ values and to compare with other authors 
plot $d^2\sigma/dq dy$ where $q= M_Z$ or $q=M_W$ respectively.

We choose the center-of-mass energy to be $\sqrt{S}=$14 TeV for the LHC 
and $\sqrt{S}=$1.96 TeV for the Tevatron.
The $Z$ boson mass is taken to be $M_Z=91.19$ GeV and the width is
$2.50$ GeV. The corresponding values for the $W$ boson are 
$M_W=80.43$ and $2.12$ GeV respectively. 
The strong coupling constant
$\alpha_s(\mu_R^2)$ is evolved using the 4-loop RG equations
depending on the order in which the cross section is evaluated.
We choose $\alpha^{\rm LO}_s(M_Z)=0.130$, $\alpha^{\rm NLO}_s(M_Z)=0.119$,
$\alpha^{{\rm N}{}^2{\rm LO}}_s(M_Z)=0.115$ and 
$\alpha^{{\rm N}{}^i{\rm LO}}_s(M_Z)=0.113$ for $i>2$.
The set MRST 2001 LO is used for leading order, MRST2001 NLO for NLO
and MRST 2002 NNLO for N${}^i$LO with $i>1$ \cite{Martin:2002dr,Martin:2001es}.
By choosing these parton densities we can compare our results with those
of other authors (see later).
We use $\alpha= 1/128$ for the electromagnetic fine structure constant, 
$\sin^2\theta_W = 0.2314$ for the weak mixing angle and
$\cos^2\theta_C = 0.975$ for the Cabibbo angle.
\begin{figure}[htb]
\vspace{1mm}
\centerline{
\epsfig{file=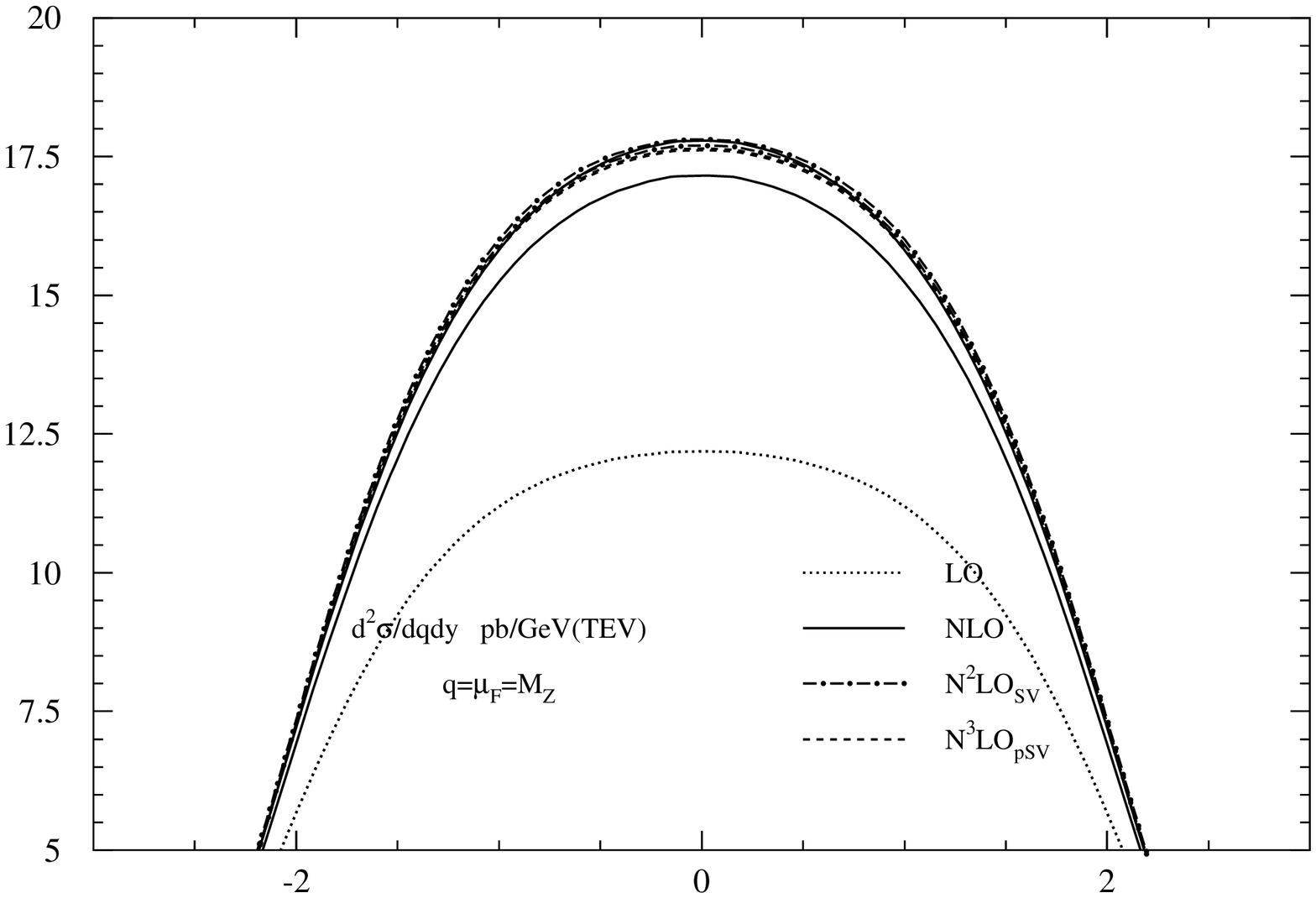,width=8cm,height=10cm,angle=0}
\epsfig{file=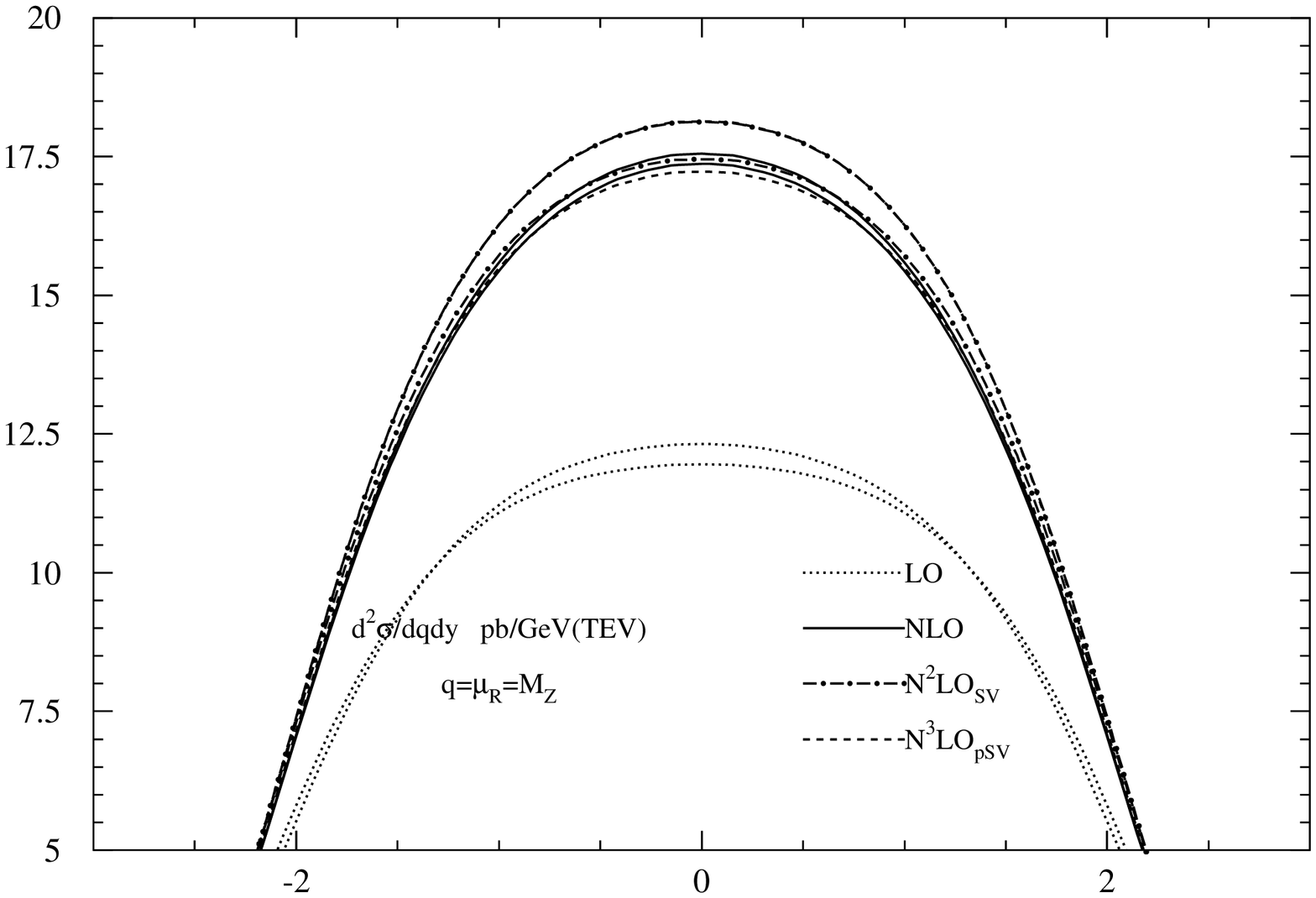,width=8cm,height=10cm,angle=0}
}
\caption{
Rapidity distributions for $Z$ boson production at the Tevatron,
and their $\mu=\mu_R$ (left panel) and $\mu=\mu_F$ (right panel) 
scale dependence (with $M^2_Z/2 < \mu^2 <2 M^2_Z$).  
The abbreviation "pSV" means partial-soft-plus-virtual.
}
\end{figure}
\begin{figure}[htb]
\vspace{1mm}
\centerline{
\epsfig{file=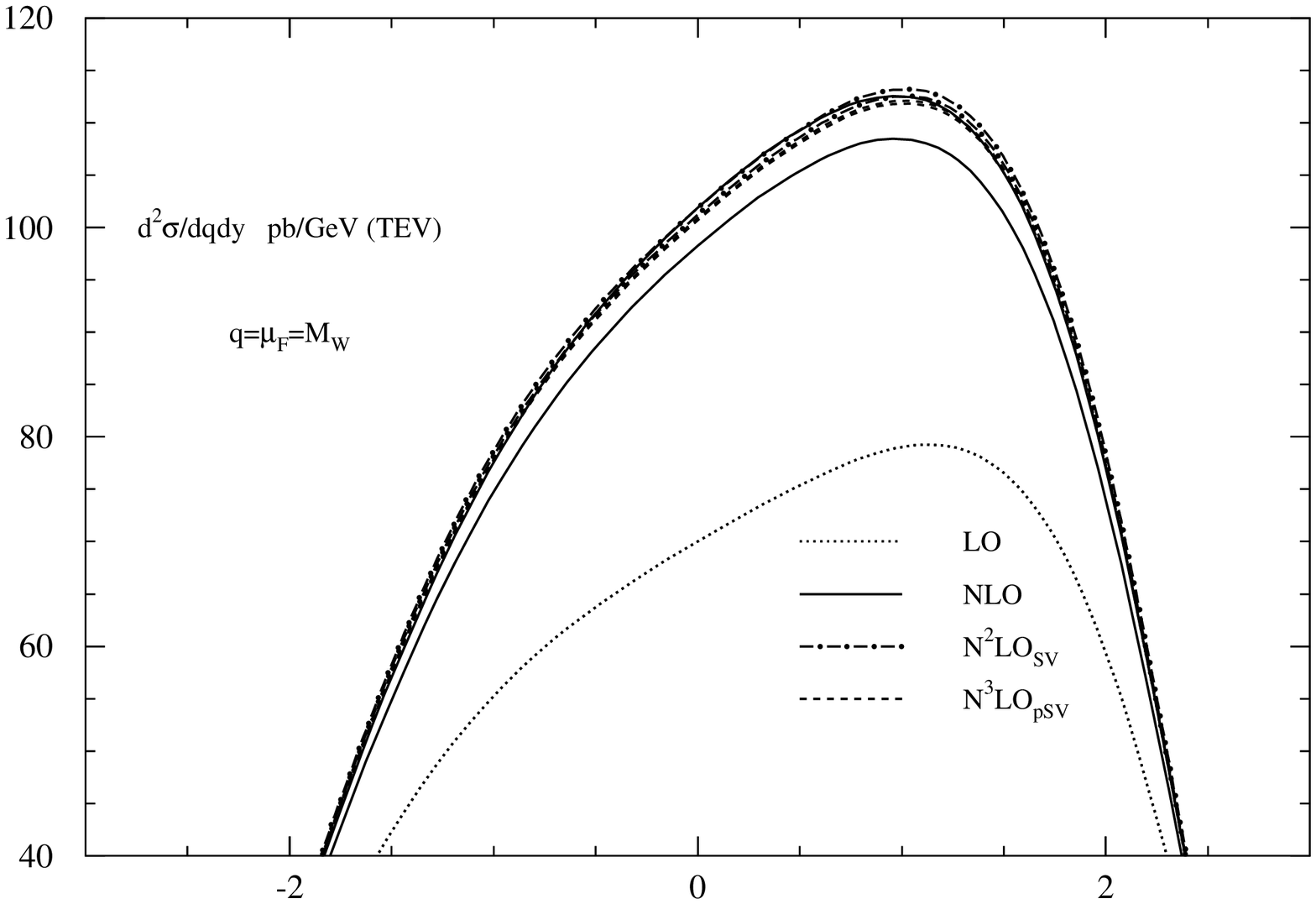,width=8cm,height=10cm,angle=0}
\epsfig{file=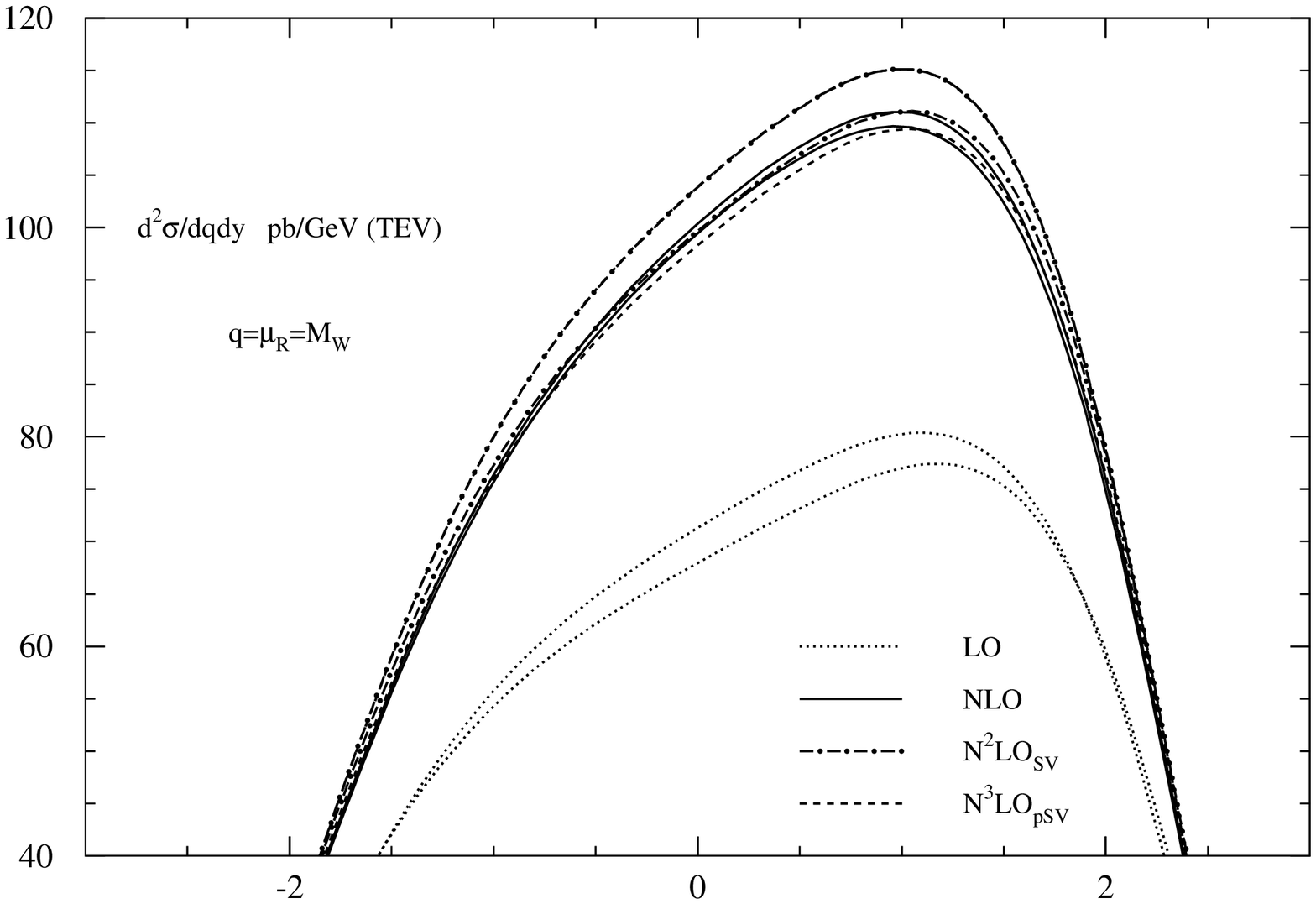,width=8cm,height=10cm,angle=0}
}
\caption{
Rapidity distributions for $W^+$ boson production at the Tevatron,
and their $\mu=\mu_R$ (left panel) and $\mu=\mu_F$ (right panel) 
scale dependence (with $M^2_W/2 < \mu^2 <2 M^2_W$).  
The abbreviation "pSV" means partial-soft-plus-virtual.
}
\end{figure}
In fig. 1 we plot the rapidity distributions for the Z-boson at the LHC 
in LO (dotted lines), NLO (solid lines), N${}^2$LO(SV only, dot-dashed lines)
and N${}^3$LO (pSV only, small-dashed lines). 
Note that we have not plotted the partial sv N${}^4$LO contributions 
and we have only plotted the curves above 20 pb/GeV to magnify the 
central rapidity region.
There are two panels in this plot and two curves in each panel
since we show the scale variations by varying the mass factorization scale 
$\mu_F$ in the parton densities and the mass renormalization scale $\mu_R$
in the coefficient functions.  
Therefore we plot the curves at fixed $\mu_F^2 = M_Z^2$ 
but with $\mu_R^2 = M_Z^2/2$ and $\mu_R^2 = 2 M_Z^2$ in the left panel 
and fixed $\mu_R^2 = M_Z^2$ 
but with $\mu_F^2 = M_Z^2/2 $ and $\mu_F^2= 2 M_Z^2$ in the right panel.
Our results in the left panel show that there is only a tiny dependence on 
the $\mu_R$ for fixed $\mu_F$ (here the LO result has no variation). 
In the right panel we see that $\mu_F$ dependence for fixed $\mu_R$
decreases as we go from LO to NLO, N${}^2$LO and N${}^3$LO respectively.
The N${}^3$LO band lies within the N${}^2$LO band and
both are within the bands for the NLO results. 
We also notice that the lower curves for the N${}^2$LO and N${}^3$LO results 
fall on top of each other. The actual numbers are different but so close 
that one cannot see this from the plot.
These results demonstrate that the perturbation series for the rapidity
distribution converges very nicely at the LHC energy. 

In fig. 2 we plot the rapidity distributions for the $W^+$-boson at the LHC 
in LO (dotted lines), NLO (solid lines), N${}^2$LO(SV only, dot-dashed lines)
and N${}^3$LO (pSV only, small-dashed lines). 
We have only plotted the curves above 200 pb/GeV to magnify the 
central rapidity region.
Again there are two panels in this plot and two curves in each panel
since we show the scale variations by varying the mass factorization scale 
$\mu_F$ in the parton densities and the mass renormalization scale $\mu_R$
in the coefficient functons.  
Therefore we plot the curves at fixed $\mu_F^2 = M_W^2$ 
but with $\mu_R^2 = M_W^2/2 $ and $ \mu_R^2= 2 M_W^2$ in the left panel 
and fixed $\mu_R^2 = M_W^2$ 
but with $\mu_F^2 = M_W^2/2$  and $\mu_F^2= 2 M_W^2$ in the right panel.
Our results in the left panel show that there is only a tiny dependence on 
the $\mu_R$ for fixed $\mu_F$ (here the LO result has no variation). 
In the right panel we see that $\mu_F$ dependence for fixed $\mu_R$
decreases as we go from LO to NLO, N${}^2$LO and N${}^3$LO respectively.
However the N${}^3$LO band is slightly below the N${}^2$LO band near $y=0$
even though both are within the bands for the NLO results,
which is probably caused by the fact that we only have a
partial soft-plus-virtual N${}^3$LO result. 
We notice again that the lower curves for the N${}^2$LO and N${}^3$LO results 
in the right panel fall on top of each other.  The actual numbers are 
different but so close that one cannot see this from the plot.
Both plots indicate that 
the perturbation series is rapidly converging at the LHC energy. 
 
In figures 3 and 4 we repeat these plots for the Tevatron energy
and the same scale choices as above.
The left panel in Fig. 3 shows excellent convergence of our results.
Note that we only plot our results above 5 pb/GeV to magnify the
central rapidity region. 
However the right panel shows that the bands for
the N${}^3$LO result are wider than those for the N${}^2$LO result and both are
wider than those for the NLO result. This can have two reasons. One is that our
results are only sv or partial sv. The other is that since the Tevatron is 
an antiproton-proton collider different combinations of
parton densities are involved as we increase the order of the 
perturbation series. We discuss this again below. 
We notice again that the lower curves for the N${}^2$LO and N${}^3$LO results 
in the right panel fall on top of each other. The actual numbers are 
different but so close that one cannot see this from the plot.

Figure 4 shows results for $W^+$ production above 40 pb/GeV
to concentrate on the central rapidity region.
The curves in the left panel show excellent convergence of the perturbation
series. The right panel again shows that the bands for the
N${}^3$LO result are wider than those for the N${}^2$LO result and both are
wider than those for the NLO result. Again we believe that this is caused by
a combination of the reasons above and comment on it below.
We notice again that the lower curves for the N${}^2$LO and N${}^3$LO results 
in the right panel fall on top of each other. The actual numbers are 
different but so close that one cannot see this from the plot.

Note that the asymmetry about $y=0$ in the rapidity plots for 
$W^+$ production at the Tevatron in Fig.4 has 
basically disappeared at the LHC energy (see Fig.2). 

We have checked our results in two ways. First by comparing our curves 
with similar plots for the rapidity distributions in \cite{Matsuura:1988sm}. 
Their computer program for the rapidity distributions 
has the exact LO result, the exact NLO result and 
a sv approximation for the N${}^2$LO result. 
We agree with their numbers when we choose their values for the electroweak 
parameters and their parton densities. Second we have also checked our 
results against those in 
\cite{Anastasiou:2003ds}, where the exact N${}^2$LO rapidity 
distributions for the $Z$ and $W^{\pm}$ bosons are calculated. Their paper
contains plots for the exact LO, the exact NLO and the exact N${}^2$LO results
in pQCD. We have run their computer code to compare their results
against ours.
Our sv approximation agrees very well with their N${}^2$LO
results in the case when $\mu_F=\mu_R =M_J/2$ for the $Z$ and $W^{\pm}$ bosons.
However we get a slightly wider band when we vary the scales in the 
N${}^2$LO case, since we only have a sv approximation. 
From this comparison we can see that the wider bands we observe in
Figures 3 and 4 are basically due to the sv and partial sv nature of our 
higher order results.
We can probably reduce the width of the bands in the N${}^3$LO case by 
calculating the missing pieces of the partial sv result.
However the central value of our partial sv N${}^3$LO result is very small 
when compared with the N${}^2$LO result indicating that the 
perturbation series continues to converge rather rapidly.
Hence this is good news for the LHC experimenters 
who plan to calibrate the Atlas and CMS detectors by measuring the rapidity and 
transverse momentum distributions of Z and $W^\pm$ bosons.
 
To summarise, we have systematically studied higher order sv
corrections to rapidity differential distributions for
$Z$ and $W^\pm$ boson production.  
We have used Sudakov resummation of soft gluons to calculate 
these processes.  The resummation of soft gluons has been achieved  
using renormalisation group invariance  
and the factorisation property of the observable that is 
considered here (the rapidity).  
Using the available information on the form factors, the DGLAP kernels and 
lower order results we have obtained compact expressions for the 
resummation of soft gluons for the rapidity distributions of $Z$ and $W^\pm$
bosons.  Using these we have computed sv rapidity distributions exactly
at N${}^2$LO and partially at N${}^3$LO.  We have presented the numerical 
impact of these results.

{\bf Acknowledgments:}  
We would like to thank Prof. P. van Baal for hospitality at the 
Lorentz Institute in Leiden where this paper was completed. 
We both acknowledge support from the FOM and the Lorentz Institute. 
Discussions with E. Laenen and A. Vogt were very helpful.
The work of J. Smith has been partially supported by the
National Science Foundation grant PHY-0098527.



\end{document}